\input{aipcheck.tex}

\documentclass{aipproc}

\usepackage{graphicx}
\usepackage{epstopdf}

\def\lesssim{\mathrel{\rlap{\lower4pt\hbox{\hskip1pt$\sim$}}}<}
\def\gtrsim{\mathrel{\rlap{\lower4pt\hbox{\hskip1pt$\sim$}}}>}

\layoutstyle{6x9}
\begin{document}

\title[Dark matter and the first stars] {Dark Stars: the First Stars in the Universe may be 
powered by Dark Matter Heating}

\author{Katherine Freese} {
address={Michigan Center for Theoretical Physics, University of Michigan, Ann Arbor, MI 48109},
,email={ktfreese@umich.edu]}}
\author{Peter Bodenheimer} {
address={Astronomy Dept., University of California, Santa Cruz, CA 95064},
,email={peter@ucolick.org}}
\author{Paolo Gondolo} {
address={Physics Dept., University of Utah, Salt Lake City, UT 84112},
,email={paolo@physics.utah.edu}}
\author{Douglas Spolyar} {
address={Physics Dept., University of California, Santa Cruz, CA 95064},
,email={dspolyar@physics.ucsc.edu}}



\begin{abstract}
A new line of research on Dark Stars is reviewed,
which suggests that the first stars to exist in the universe
were powered by dark matter heating rather than by fusion.
Weakly Interacting Massive Particles, which may be there own
antipartmers, collect inside the first stars and annihilate to produce
a heat source that can power the stars.  A new stellar phase results,
a Dark Star, powered by dark matter annihilation as long as there is
dark matter fuel.
\end{abstract}

\date{\today}

\maketitle

\section{Introduction}

The first stars to form in the universe, at redshifts $z \sim 10-50$,
may be powered by dark matter annihilation for a significant period of
time  \cite{sfg}.  We have dubbed these objects ``Dark Stars.''

Weakly Interacting Massive Particles (WIMPs)
 are the best motivated dark matter candidates.  WIMP annihilation in the
early universe provides the right abundance today to explain the dark
matter content of our universe. This same annihilation process will
take place at later epochs in the universe wherever the dark matter
density is sufficiently high to provide rapid annihilation.  The first
stars to form in the universe are a natural place to look for
significant amounts of dark matter annihilation, because they form at
the right place and the right time. They form at high redshifts, when
the universe was still substantially denser than it is today, and at
the high density centers of dark matter haloes. 
 
The first stars form inside dark matter (DM) haloes of $10^6 M_\odot$
(for reviews see e.g. \cite{Ripamonti:2005ri, Barkana:2000fd, Bromm:2003vv, 
Yoshida:2008gn}; see also \cite{ABN, Yoshida06}.)
One star is thought to form inside one such DM halo. The first stars may
play an important role in reionization, in seeding supermassive black
holes, and in beginning the process of production of heavy elements in
later generations of stars.  

It was our idea to ask, what is the
effect of the DM on these first stars?  We studied the behavior of
WIMPs in the first stars, and  found that they 
can radically alter the stellar evolution.  The annihilation products of the
dark matter inside the star can be trapped and
deposit enough energy to heat the star and prevent it from further
collapse.  A new stellar phase results, a Dark Star, powered
by DM annihilation as long as there is DM fuel, for millions to billions of years.

\subsection{Weakly Interacting Dark Matter}

\label{sec:WIMPs}

WIMPs are  natural dark matter candidates from particle physics.
These particles, if present in thermal abundances in the early
universe, annihilate with one another so that a predictable number of
them remain today.  The relic density of these particles is
\begin{equation}
\Omega_\chi h^2 = (3 \times 10^{-26} {\rm cm}^3/{\rm sec})
/ \langle \sigma v \rangle_{ann}
\end{equation}
where the annihilation cross section $\langle \sigma v \rangle_{ann} $
of weak interaction strength automatically gives the right answer, near the
WMAP  \cite{Komatsu:2008hk} value $\sim 23\%$.
This coincidence is known as "the WIMP miracle" and is the reason why
WIMPs are taken so seriously as DM candidates.  The best WIMP
candidate is motivated by Supersymmetry (SUSY): the lightest
neutralino in the Minimal Supersymmetric Standard Model
(see the reviews by \cite{jkg, jkgb, jkgc, Bertone_etal2004}).

This same annihilation process is also the basis for
DM indirect detection searches.  The first paper discussing annihilation in stars was \cite{Krauss:1985ks}; the first papers suggesting searches for annihilation products of WIMPs in the Sun
were by Silk {\it et al} \cite{SOS}; and in the Earth
by Freese \cite{freese} as well as Krauss, Srednicki and
Wilczek \cite{ksw}.  Other studies of WIMPs in today's stars include \cite{salati, Moskalenko:2007ak, Scott:2008uw, Scott:2007md}.
This talk reviews the study of WIMP annihilation as a heat source for the first stars.

As our canonical parameter values, we take $m_\chi =
100$GeV for the WIMP mass and $\langle \sigma v \rangle_{ann} = 3
\times 10^{-26} {\rm cm^3/sec}$ for the annihilation cross section
but consider a variety of masses and cross sections.

\section{Three Criteria for Dark Matter Heating}

 WIMP annihilation produces energy at a rate per
unit volume 
\begin{equation}
\label{eq:heat}
  Q_{\rm ann} = \langle \sigma v \rangle_{ann} \rho_\chi^2/m_\chi
  \linebreak \simeq  10^{-29} {{\rm erg} \over {\rm cm^3/s}} \,\,\, {\langle
    \sigma v \rangle \over (3 \times 10^{-26} {\rm cm^3/s})} \left({n \over {\rm
        cm^{-3}}}\right)^{1.6} \left({100 {\rm GeV}\over m_\chi}\right) 
\end{equation}
where $\rho_\chi$ is the DM energy density inside the star and $n$ is
the stellar hydrogen density.  Paper I \cite{sfg} outlined the three key ingredients
for Dark Stars: 1) high dark matter densities, 2) the annihilation
products get stuck inside the star, and 3) DM heating wins over other
cooling or heating mechanisms.  These same ingredients are required
throughout the evolution of the dark stars, whether during the
protostellar phase or during the main sequence phase.

{\bf First criterion: High Dark Matter density inside the star.}  One can see
from Eq.(\ref{eq:heat}) that the DM annihilation  rate scales as WIMP density squared,
because two WIMPs must find each other to annihilate.  Thus the annihilation is
significant wherever the density is high enough.  Dark
matter annihilation is a powerful energy source in these first stars (and not in today's stars)
because the dark matter density is high.  First, DM densities in the early universe were
higher by $(1+z)^3$.   Second, the first stars form exactly in the centers of DM haloes
where the densities are high (as opposed to today's stars which are scattered throughout
the disk of the galaxy rather than at the Galactic Center).  We assume for
our standard case that the DM density inside the $10^6 M_\odot$ DM halo initially has an 
NFW (Navarro, Frenk \& White \cite{NFW})
 profile for both DM and gas, which has
substantial DM in the center of the halo. Third, a further DM enhancement
takes place in the center of the halo: as the protostar forms, it deepens the potential
well at the center and pulls in more DM as well.  We have computed this
enhancement in several ways \cite{sfg} (see also \cite{Natarajan:2008db}); 
most recently we performed an exact calculation \cite{Freese:2008hb}.
Fourth, at later stages, we also consider possible further enhancement
due to capture of DM into the star (discussed below).

{\bf Second Criterion: Dark Matter Annihilation Products get stuck
  inside the star}.  In the early stages of Pop III star formation,
when the gas density is low, most of the annihilation energy is
radiated away \cite{Ripamonti:2006gr}.  However, as the gas collapses
and its density increases, a substantial fraction $f_Q$ of the
annihilation energy is deposited into the gas, heating it up at a rate
$f_Q Q_{\rm ann}$ per unit volume.  While neutrinos escape from the
cloud without depositing an appreciable amount of energy, electrons
and photons can transmit energy to the core.  We have computed
estimates of this fraction $f_Q$ as the core becomes more dense. Once
$n\sim 10^{11} {\rm cm}^{-3}$ (for 100 GeV WIMPs), e$^-$ and photons
are trapped and we can take $f_Q \sim 2/3$.

{\bf Third Criterion: DM Heating is the dominant heating/cooling
  mechanism in the star}.  We find that, for WIMP mass $m_\chi =
100$GeV (1 GeV), a crucial transition takes place when the gas density
reaches $n> 10^{13} {\rm cm}^{-3}$ ($n>10^9 {\rm cm}^{-3}$).  Above
this density, DM heating dominates over all relevant cooling
mechanisms, the most important being H$_2$ cooling \cite{Hollenbach}.

Figure 5 shows evolutionary tracks of the protostar in the
temperature-density phase plane with DM heating included
(Yoshida et al. \cite{Yoshida_etal08}), for two DM particle
masses (10 GeV and 100 GeV).  Moving to the right on this plot is
equivalent to moving forward in time.  Once the black dots are
reached, DM heating dominates over cooling inside the star, and the
Dark Star phase begins.  The protostellar core is prevented from
cooling and collapsing further.  The size of the core at this point is
$\sim 17$ A.U. and its mass is $\sim 0.6 M_\odot$ for 100 GeV mass
WIMPs.  A new type of object is created, a Dark Star supported by DM
annihilation rather than fusion.

\begin{figure}
  \includegraphics[height=.3\textheight]{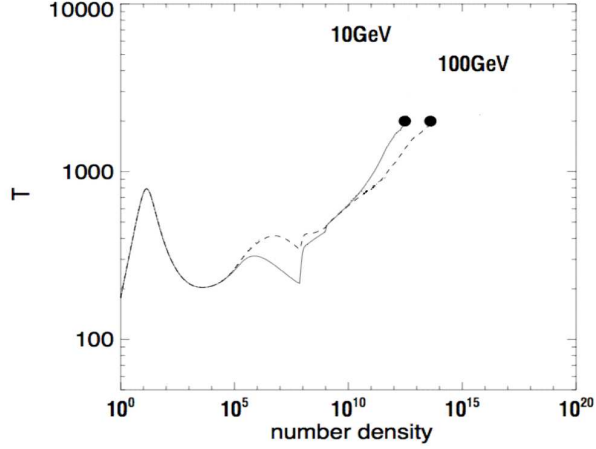}
\caption{ Temperature (in degrees K) as a function of hydrogen density
  (in cm$^{-3}$) for the first protostars, with DM annihilation
  included, for two different DM particle masses (10 GeV and 100 GeV).
  Moving to the right in the figure corresponds to moving forward in
  time.  Once the ``dots'' are reached, DM annihilation wins over H2
  cooling, and a Dark Star is created.}
\end{figure}

\section{Building up the Mass}

We have found the stellar structure of the dark stars
(hereafter DS) \cite{Freese:2008wh}.  They accrete mass from the
surrounding medium.  In our paper we build up the DS mass as it grows
from $\sim 1 M_\odot$ to $\sim 1000 M_\odot$.
As the mass increases, the DS radius adjusts
until the DM heating matches its radiated luminosity.  We find
polytropic solutions for dark stars in hydrostatic and thermal
equilibrium. We build up the DS by accreting $1 M_\odot$ at a time
with an accretion rate of $2 \times 10^{-3} M_\odot$/yr, always
finding equilibrium solutions.  We find that initially the DS are in
convective equilibrium; from $(100-400) M_\odot$ there is a transition
to radiative; and heavier DS are radiative.  As the DS grows, it pulls
in more DM, which then annihilates.  We continue this process until
the DM fuel runs out at $M_{DS} \sim 800 M_\odot$ (for 100 GeV WIMPs).
Figure 6 shows the stellar structure. 

One can see ``the power of
darkness:'' although the DM constitutes a tiny fraction ($<10^{-3}$)
of the mass of the DS, it can power the star. The reason is that WIMP
annihilation is a very efficient power source: 2/3 of the initial
energy of the WIMPs is converted into useful energy for the star,
whereas only 1\% of baryonic rest mass energy is useful to a star via
fusion.

\begin{figure}
 \includegraphics[height=.3\textheight]{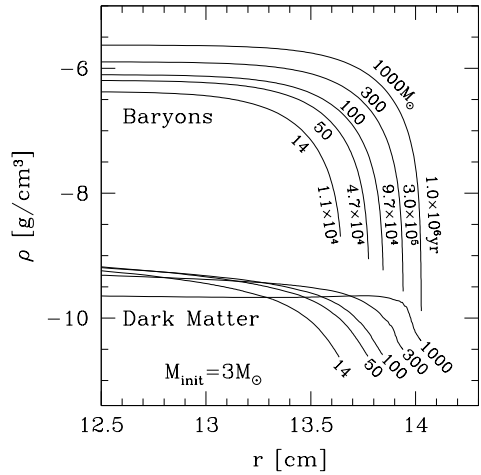}
\caption{Evolution of a dark star (n=1.5) as mass is accreted onto the
  initial protostellar core of 3 M$_\odot$.  The set of upper (lower)
  curves correspond to the baryonic (DM) density profile at different
  masses and times. Note that DM constitutes $<10^{-3}$ of the mass of
  the DS.}
\end{figure}

\section{Results and Predictions}

Our final result \cite{Freese:2008wh} is very large first
  stars; e.g., for 100 GeV WIMPs, the first stars have $M_{DS} = 800
M_\odot$.  Once the DM fuel runs out inside the DS, the star contracts
until it reaches $10^8$K and fusion sets in.  A possible end result of
stellar evolution will be large black holes.  The Pair
Instability SN \cite{HW} that would be produced from 140-260 $M_\odot$ stars (and whose chemical
imprint is not seen) would not be as abundant.  Indeed this process
may help to explain the supermassive black holes that have been found
at high redshift ($10^9 M_\odot$ BH at z=6) and are, as yet,
unexplained \cite{li, Pelupessy}.  The stars are very bright,
$\sim 10^6 L_\odot$, and relatively cool, (6000-10,000)K (as opposed
to standard Pop III stars whose surface temperatures exceed $30,000K$).  
One can thus
hope to find DS and differentiate them from standard Pop III stars.

\section{Later stages: Capture}

The dark stars will last as long as the DM fuel inside them persists.  The
original DM inside the stars runs out in about a million years.
However, as discussed in the next paragraph, the DM may be replenished
by capture, so that the DS can live indefinitely due to DS
annihilation.  We suspect that the DS will eventually leave their high
density homes in the centers of DM haloes, especially once mergers of
haloes with other objects takes place, and then the DM fuel will run
out. The star will eventually be powered by fusion. Whenever it again
encounters a high DM density region, the DS can capture more DM and be
born again.

The new source of DM in the first stars is capture of DM particles
from the ambient medium.  Any DM particle that passes through the
DS has some probability of interacting with a nucleus in the star
and being captured. The new particle physics ingredient required
here is a significant scattering cross section between the WIMPs
and nuclei. Whereas the annihilation cross section is
fixed by the relic density, the scattering cross section is a
somewhat free parameter, set only by bounds from direct detection 
experiments.  
Two simultaneous papers  \cite{Freese:2008ur, Iocco:2008xb} found the
same basic idea: the DM luminosity from captured WIMPs can be larger
than fusion for the DS. Two uncertainties exist here: the scattering
cross section, and the amount of DM in the ambient medium to capture
from.  DS studies following the original papers that include
capture have assumed (i)  the maximal scattering cross sections allowed by
experimental bounds and (ii) ambient DM densities that are never depleted.
With these assumptions, DS evolution models with DM heating after the
onset of fusion have now been studied in several papers
\cite{Iocco:2008rb, Taoso:2008kw, Yoon:2008km}.
 Studies of effects on reionization have been
looked at by \cite{Schleicher:2008gk}.

In short, the first stars to form in the universe may be Dark Stars
powered by DM heating rather than by fusion.  Our work indicates that
they may be very large ($800 M_\odot$ for 100 GeV mass WIMPs). Once DS
are found, one can use them as a tool to study the properties of WIMPs.

\section{Conclusion}
95\% of the mass in galaxies and clusters of galaxies is in the form
of an unknown type of dark matter.  One of the key properties of WIMP candidates is its
annihilation cross section, yielding the proper relic density
today. As a consequence of this annihilation, the first stars in the
universe may provide another avenue to test the DM hypothesis. These
stars may be powered by DM annihilation, and one can look for them in
upcoming telescopes.  It is an exciting prospect to discover a new type of
star powered by the dark matter in the universe.

\begin{theacknowledgments} K. Freese thanks her collaborators in this research: Anthony Aguirre, Peter Bodenheimer, Paolo Gondolo, and Doug Spolyar. She also thanks Naoki Yoshida for Figure 1. She ackhowledges support from the DOE and MCTP via the University of Michigan.
\end{theacknowledgments}


\begin{thebibliography}{99}

\bibitem{sfg}
  D. Spolyar, K.~Freese, and P.~Gondolo,
  astro-ph/0705.0521.



\bibitem{Ripamonti:2005ri}
  E.~Ripamonti and T.~Abel,
  astro-ph/0507130.

\bibitem{Barkana:2000fd}
  R.~Barkana and A.~Loeb,
  Phys.\ Rept.\  {\bf 349}, 125 (2001).

\bibitem{Bromm:2003vv}
  V.~Bromm and R.~B.~Larson,
  Ann.\ Rev.\ Astron.\ Astrophys.\  {\bf 42}, 79 (2004).
  
  \bibitem{Yoshida:2008gn}
  N.~Yoshida, K.~Omukai and L.~Hernquist,
Science 321 (2008) 669-671,

\bibitem{ABN}
  T.~Abel, G.~L.~Bryan and M.~L.~Norman,
  Science {\bf 295}, 93 (2002).
  
  \bibitem{Yoshida06}
  N.~Yoshida et al., 
  Astrophys.\ J.\  {\bf 652}, 6 (2006).
  
  \bibitem{Komatsu:2008hk}
  E.~Komatsu {\it et al.}  [WMAP Collaboration],
  arXiv:0803.0547 [astro-ph].
  
  \bibitem{jkg}
Jungman, G., Kamionkowski, M., \&  Griest, K.,
  Phys.\ Rept.,   {\bf 267}, 195 (1996).

\bibitem{jkgb} Lewin and P. \& Smith, Astropart.\ Phys.\ {\bf 6} 87-112 (1996)

\bibitem{jkgc} J. Primack, D. Seckel, and B. Sadoulet, Ann. Rev.
  of Part. and Nucl. Science, {\bf 38} 751 (1988).

\bibitem{Bertone_etal2004}
Bertone, G., Hooper, D. \& Silk, J.,
Phys.\ Rept.\  {\bf 405}, 279 (2005)

\bibitem{Krauss:1985ks}
  L.~M.~Krauss, K.~Freese, W.~Press and D.~Spergel,
  Astrophys.\ J.\  {\bf 299}, 1001 (1985).

\bibitem{SOS}
  M. Srednicki, K.A. Olive, and J. Silk, Nucl.\ Phys.\ B {\bf 279},
  804 (1987).

\bibitem{freese}
  K.~Freese,
  Phys.\ Lett.\  B {\bf 167}, 295 (1986).

\bibitem{ksw} 
  L.M. Krauss, M. Srednicki, and F. Wilczek, Phys.\ Rev.\ D {\bf 33},
  2079 (1986).
  
  \bibitem{salati}
P. Salati  \& J. Silk,  ApJ, {\bf 338}, 24 (1989).

\bibitem{Moskalenko:2007ak}
  I.~V.~Moskalenko and L.~L.~Wai,
  Astrophys.\ J.\  {\bf 659}, L29 (2007)
  [arXiv:astro-ph/0702654].

  
\bibitem{Scott:2008uw}
  P.~Scott, M.~Fairbairn and J.~Edsjo,
  arXiv:0810.5560 [astro-ph].
  
\bibitem{Scott:2007md}
  P.~Scott, J.~Edsjo and M.~Fairbairn,
  arXiv:0711.0991 [astro-ph].
  
\bibitem{NFW}
  J.~F.~Navarro, C.~S.~Frenk and S.~D.~M.~White,
  Astrophys.\ J.\  {\bf 462}, 563 (1996).

\bibitem{Natarajan:2008db}
  A.~Natarajan, J.~C.~Tan and B.~W.~O'Shea,
  arXiv:0807.3769 [astro-ph].

\bibitem{Freese:2008hb}
  K.~Freese, P.~Gondolo, J.~A.~Sellwood and D.~Spolyar,
  arXiv:0805.3540 [astro-ph].
  
  \bibitem{Ripamonti:2006gr}
  E.~Ripamonti, M.~Mapelli and A.~Ferrara,
  Mon.\ Not.\ Roy.\ Astron.\ Soc.\  {\bf 375}, 1399 (2007).

\bibitem{Hollenbach}
  D.~Hollenbach and C.~F.~McKee,
   Astrophys.\ J.\  Suppl.\ {\bf 41}, 555 (1979).
   
\bibitem{Yoshida_etal08}
N. Yoshida, K. Freese, P. Gondolo, and D.Spolyar, work in preparation.

\bibitem{Freese:2008wh}
  K.~Freese, P.~Bodenheimer, D.~Spolyar and P.~Gondolo,
  arXiv:0806.0617 [astro-ph].


\bibitem{mckee} J.~C.~Tan and C.~F.~McKee,
  Astrophys.\ J.\  {\bf 603}, 383 (2004).
  
  \bibitem{HW}
A. Heger \&  S.E.Woosley, 
  ApJ {\bf 567}, 532  (2002)
  
\bibitem{li}
  Y.~X.~Li {\it et al.},
  arXiv:astro-ph/0608190.

\bibitem{Pelupessy}
  F.~I.~Pelupessy, T.~Di Matteo and B.~Ciardi,
  arXiv:astro-ph/0703773.
  
 
\bibitem{Freese:2008ur}
  K.~Freese, D.~Spolyar and A.~Aguirre,
  JCAP {\bf 0811}, 014 (2008)
  [arXiv:0802.1724 [astro-ph]].
  
  \bibitem{Iocco:2008xb}
  F.~Iocco,
  Astrophys.\ J.\  {\bf 677}, L1 (2008)
  [arXiv:0802.0941 [astro-ph]].
  
  \bibitem{Iocco:2008rb}
  F.~Iocco, A.~Bressan, E.~Ripamonti, R.~Schneider, A.~Ferrara and P.~Marigo,
  arXiv:0805.4016 [astro-ph].
  
  \bibitem{Taoso:2008kw}
  M.~Taoso, G.~Bertone, G.~Meynet and S.~Ekstrom,
  arXiv:0806.2681 [astro-ph].
  
  \bibitem{Yoon:2008km}
  S.~C.~Yoon, F.~Iocco and S.~Akiyama,
  arXiv:0806.2662 [astro-ph].

\bibitem{Schleicher:2008gk}
  D.~R.~G.~Schleicher, R.~Banerjee and R.~S.~Klessen,
  arXiv:0809.1519 [astro-ph].

\end{thebibliography}
\end{document}